\documentclass[prd,twocolumn,showpacs,amsmath,amssymb,widetext,floatfix]{revtex4-1}
\usepackage{graphicx}% Include figure files
\usepackage{dcolumn}% Align table columns on decimal point
\usepackage{bm}% bold math
\usepackage{epsfig}
\usepackage{color}
\usepackage{hyperref}

\hypersetup{
    colorlinks=true,
    linkcolor=red,
    citecolor=blue,
}

\def\fun#1#2{\lower3.6pt\vbox{\baselineskip0pt\lineskip.9pt
  \ialign{$\mathsurround=0pt#1\hfil##\hfil$\crcr#2\crcr\sim\crcr}}}
\def\simgt{\mathrel{\lower0.6ex\hbox{$\buildrel {\textstyle >}
 \over {\scriptstyle \sim}$}}}
\def\simlt{\mathrel{\lower0.6ex\hbox{$\buildrel {\textstyle <}
 \over {\scriptstyle \sim}$}}}

\input epsf

\newcommand{\mnras}{MNRAS}
\newcommand{\apjl}{ApJL}
\newcommand{\aj}{AJL}

\newcommand{\gth}{G_\Theta} 
\newcommand{\scut}{\sigma_{\rm cut}} 

\def\be{\begin{equation}}
\def\ee{\end{equation}}
\def\ba{\begin{eqnarray}}
\def\ea{\end{eqnarray}}

\newcommand{\hompc}{\,h\,{\rm Mpc}^{-1}}
\newcommand{\mpcoh}{\,h^{-1}\,{\rm Mpc}}

\newcommand{\zeff}{z_{\rm eff}} 
 
\newcommand{\jcap}{J. Cosmology Astropart. Phys.}

\begin{document}

\preprint{}

%\title{Anisotropic Analysis on Redshift Space Clustering using BOSS DR11}
%\title{Confirming and Questioning Standard LCDM model at local universe using BOSS DR11} 
\title{Cosmological Tests using Redshift Space Clustering in BOSS DR11}
 
\author{Yong-Seon Song$^{1}$, Cristiano G.\ Sabiu$^2$, Teppei Okumura$^3$, Minji Oh$^{1,4}$, Eric V.\ Linder$^{1,5}$} 
\email{ysong@kasi.re.kr}
\affiliation{$^1$Korea Astronomy and Space Science Institute, Daejeon 305-348, 
Korea} 
\affiliation{$^2$Korea Institute for Advanced Study, Dongdaemun-gu, Seoul 
130-722, Korea}   
\affiliation{$^3$Kavli Institute for the Physics and Mathematics of the Universe (Kavli IPMU, WPI), The University of Tokyo, Chiba 277-8582, Japan} 
\affiliation{$^4$University of Science and Technology, Daejeon 305-333, Korea}
\affiliation{$^5$Berkeley Lab and Berkeley Center for Cosmological Physics, 
University of California, Berkeley, CA 94720, USA}
\date{\today}

\begin{abstract}
We analyze the clustering of large scale structure in the Universe in a model independent method, accounting for anisotropic effects along and transverse to the line of sight. The Baryon Oscillation Spectroscopy Survey Data Release 11 provides a large sample of 690,000 galaxies, allowing determination of the Hubble expansion $H$, angular distance $D_A$, and growth rate $\gth$ at an effective redshift of $z=0.57$. After careful bias and convergence studies of the effects from small scale clustering, we find that cutting transverse separations below 40 Mpc/$h$ delivers robust results while smaller scale data leads to a bias due to unmodelled nonlinear and velocity effects. The converged results are in agreement with concordance $\Lambda$CDM cosmology, general relativity, and minimal neutrino mass, all within the 68\% confidence level. We also present results separately for the northern and southern hemisphere sky, finding a slight tension in the growth rate -- potentially a signature of anisotropic stress, or just covariance with small scale velocities -- 
but within 68\% CL. 
\end{abstract}



\maketitle

\section{Introduction}
Three dimensional maps of galaxy positions over wide sky areas are 
greatly advancing our cosmological knowledge. The clustering of galaxies 
measures the growth of large scale structure and echoes the baryon-photon 
sound horizon scale in the extra power of the baryon acoustic oscillation 
feature. Studying clustering along the line of sight (in the redshift 
direction) and transverse to the line of sight (in the angular direction) 
probes the Hubble expansion and the angular diameter distance 
respectively~\cite{1996MNRAS.282..877B,1996ApJ...470L...1M, Hu:2003ti,Padmanabhan:2008ag}. 
Such anisotropic effects have been extensively analyzed in various redshift surveys \cite[e.g.,][]{2008ApJ...676..889O,Gaztanaga:2008xz,Reid:2012sw,2013arXiv1312.4611B,2014MNRAS.439.3504S,2014MNRAS.440.2692S}. Combining the radial and 
transverse information can measure the cosmically induced shear of the 
clustering, known as the Alcock-Paczynski effect~\cite{Alcock:1979mp,corredoira}. 

Here we use the 690,000 galaxies of the Baryon Oscillation Spectroscopic 
Survey (BOSS) Data Release 11 (DR11) to carry out a model independent 
anisotropic clustering analysis, measuring the Hubble parameter $H$, 
angular distance $D_A$, and growth rate variable $\gth$ in a volume with 
effective redshift $\zeff=0.57$. This analysis does not assume any specific 
dark energy model or even the Friedmann-Robertson-Walker relation between 
the expansion rate $H$ and distance $D_A$, nor the general relativity 
relation between expansion and growth $\gth$~\cite{Song:2013ejh}.

This work closely follows our approach~\cite{Song:2013ejh,Linder:2013lza} with BOSS DR9 simulation and data, with 
several improvements arising from both the BOSS data 
(see \cite{Linder:2013lza}) and our analysis. The data covers 
a wider sky area, much more uniformly, and the computational simulations for 
the mock catalogs take into account more instrumental effects. In our 
analysis the improved data quality allows straightforward use of the 
covariance matrix without need for the previous singular value decomposition 
to control noise. We also study in more detail the dependence of the results 
on the small scale clustering and their convergence behavior as this regime, 
with uncertain nonlinearity and velocity effects, is truncated. 

Section~\ref{sec:theory} summarizes briefly our approach to fitting the 
clustering correlation function, concentrating on differences from 
\cite{Song:2013ejh,Linder:2013lza}. 
We measure the anisotropic clustering in Section~\ref{sec:measure}. In Section~\ref{sec:results} we present the joint likelihood results for the
cosmological quantities of the expansion $H$, distance $D_A$,
and growth $\gth$, and assess consistency with the concordance cosmology, neutrino mass, and general relativity. We also investigate the dependence of the
cosmological results on the small scale cut--off, in terms of
both bias and precision, and explore the comparison of northern and southern Galactic hemisphere data. We summarize and conclude in Section~\ref{sec:concl}.

%%%%%%%%%%%%%%%%%%%%%%%%%%%%%%%%%%%%%%%%%%%%
\section{Theoretical Model and Simulation} \label{sec:theory} 

%%%%%%%%%%%%%%%%%%%%%%%%%%%%%%%%
\subsection{Theoretical model}

The observed galaxy two--point correlation function, $\xi$, is given by 
\ba\label{eq:xi_eq}
\xi(\sigma,\pi)&=&\int \frac{d^3k}{(2\pi)^3} {P}(k,\mu)e^{i{\bf k}\cdot{\bf s}} \ , 
\ea
where $\sigma$ and $\pi$ are the 
separations between the galaxies of the pair in the transverse and 
radial directions respectively, with respect to the observer, and $s$ is 
the total separation $s=(\sigma^2+\pi^2)^{1/2}$. The correlation function 
is the Fourier transform of the power spectrum $P$, with $\vec k$ the 
wavemode and $\mu$ the cosine of the angle between $\vec k$ and the line 
of sight. 

The two point correlation function $\xi$ as observed in redshift space 
(i.e.\ where the radial dimension is not true separation but distance as 
measured through redshift, involving both separation and radial velocity) 
involves two distinct effects from redshift space distortions (RSD) \cite{Kaiser:1987qv,1998ASSL..231..185H}. 
In the linear regime, the density fluctuations and peculiar velocities are coherently evolved through the continuity equation. Thus the known correlation function in real space from the linear perturbation theory developed by gravitational instability is uniquely transformed into $\xi(\sigma,\pi)$ in redshift space. 
Gravitational infall squeezes the clustering pattern in redshift space 
along the line of sight (i.e.\ the $\pi$-direction), enhancing the 
correlation function by the Kaiser factor \cite{Kaiser:1987qv}.  In the 
non--linear regime, however, the observed correlation function appears 
elongated along the line of sight due to the random virial velocities of 
galaxies, called the Finger of God effect (FoG) \cite{1972MNRAS.156P...1J}. Because the FoG effect 
smears into even large scales, the linear theory Kaiser effect is not 
appropriate along the $\pi$ direction~\cite{1999ApJ...517..531S, Scoccimarro:2004tg}. 

A more accurate description of these RSD effects is required for BOSS 
clustering data. Improved models were tested using simulations 
in~\cite{Song:2013ejh}, and applied to DR9 data in~\cite{Linder:2013lza}. 
These corrections are  briefly reviewed below. 

The mapping between real space and redshift space has significant 
correlations between the density and velocity fields. Although it extends 
into a infinite series of polynomials, a few leading modes are dominant 
near the linear regime. While the original linear theory includes terms 
up to $\mu^4$, terms up to $\mu^6$ are necessary in the improved models; 
higher order terms can be safely ignored in the quasi-linear regime~\cite{Taruya:2010mx}.

The cross--correlation spectrum between density $\delta$ and velocity 
$\Theta$ is not independently measured, but rather estimated from the measured 
auto--correlations of $\delta$ and $\Theta$, based upon the assumption of a 
perfect cross--correlation coefficient (which is valid only in the linear 
theory). The deviation of spectra from the linear theory is perturbatively calculated using the resummed perturbation theory called {\tt RegPT}~\cite{Taruya:2007xy,Taruya:2012ut}. When restricting analysis to the quasi-linear regime, the result is the non--linear portions of the power spectra are better separated from the linear spectra; for the latter the assumption of perfect cross--correlation between density and velocity fields can be applied. 

However, the FoG effect remains non--perturbative in this model. The 
non--linear smearing effect is dominant at first order, and can be 
parameterised by a velocity dispersion $\sigma_p$. Our theoretical models 
(as all others) are broken at scales in which higher order terms become 
important, however. 
This motivates us to introduce the cut--off scales  and consider them 
carefully. We use two cut--offs: $s_{\rm cut}$ and $\sigma_{\rm cut}$. 
The $s_{\rm cut}$ accounts for the limit of theoretical description of 
{\tt RegPT}, and the $\sigma_{\rm cut}$ reflects the unknown contamination of the residual FoG effect. 

%%%%%%%%%%%%%%%%%%%%%%%%%%%%%%%% 
\subsection{Methodology and simulation test}

The observed clustering as a function of the transverse and radial distances 
is related to the density and velocity growth functions, and the FoG 
parameter $\sigma_p$. Cosmological information is extracted from the density 
and velocity functions. 

From the clustering $\xi(\sigma,\pi)$, measured in comoving distances, the 
transverse and radial distances scale linearly with $D_A$ and $H^{-1}$ 
respectively. 
The evolution of clustering occurs coherently for all scales in linear 
theory, and its initial shape (scale dependence) is determined in the early 
universe. Using early universe information from the cosmic microwave 
background we denote this as a Planck (or WMAP9) prior. All evolution in 
the amplitude after the last scattering epoch informs us about late time 
cosmology. We denote the normalized density and coherent motion (velocity) 
growth functions as $G_b$ and $G_{\Theta}$. This is a model independent 
analysis in the sense that we do not require, or use, any specific 
assumptions on energy density components such as dark energy or curvature; 
indeed we do not even have to assume the Friedmann-Robertson-Walker relation 
of $D_A$ as an integral of $H^{-1}$. Note that 
$G_b=b G_\delta$, where $b$ is the galaxy bias and in linear theory 
$G_\delta=D$, where $D$ is the growth factor. Similarly $\gth=dD/d\ln a$ is 
growth rate, and is proportional to the sometimes used combination 
$f\sigma_8$~\cite{Song:2008qt}, with $\gth=f\sigma_8\,(D_0/\sigma_{8,0})$ where 0 denotes the 
present. See \cite{Linder:2013lza} for more details. 

The spectra of the density and the velocity fields are naturally expected to receive nonlinear corrections. One of these corrections comes from the 
random motion of galaxies, which results in the damping effect of the power spectrum amplitude. We apply a Gaussian FoG function with free parameter 
$\sigma_p$ characterizing the velocity. However, as these non--perturbative 
damping effects are not fully understood, we employ a cut-off scale to 
remove small scales where this is exacerbated, and study the effects of 
varying that scale. 

In order to check the validity of our overall approach, we test it against 
simulations. 
We use the mock galaxy catalogs created by~\cite{Manera:2012sc}, which are designed to investigate the various systematics in the galaxy sample from Data Release 11 (DR11) of the Baryon Oscillation Spectroscopic Survey (BOSS) \citep{Schlegel:2009,Eisenstein:2011,2012MNRAS.427.3435A}, referred to as the ``CMASS" galaxy sample. In constructing the mock galaxy catalogs,~\cite{Manera:2012sc} utilized second-order Lagrangian perturbation theory (2LPT) for the galaxy clustering driven by gravity, which enables the creation of a mock catalogs much faster than running an $N$-body simulation. The redshift range of galaxies in the catalog is $0.43<z<0.7$ and each catalog contains $\sim 7 \times 10^5$ galaxies, 90\% of which are central galaxies residing in dark matter halos of $\sim 10^{13}h^{-1}M_\odot$. Table~\ref{table:simulation} illustrates that our analysis successfully 
recovers the simulated values for the cosmological quantities 
$D_A$, $H^{-1}$ and $G_{\Theta}$. Our studies show that 
$s_{\rm cut}=50 \mpcoh$ and $\sigma_{\rm cut}=40 \mpcoh$ give converged, 
robust results. 

%%%%%%%%%%%%%%%%%%%%%%%%%%%%%
\begin{table}
\begin{center}
\begin{tabular}{lcc}
\hline
\hline
Parameters & \quad Simulated values\quad  & \quad Measured values\quad \\
\hline
$D_A\,(\mpcoh)$       & $932.6$ & $939.7^{+26.7}_{-32.6}$  \\
$H^{-1}\,(\mpcoh)$   & $2177.5$ & $2120.5^{+82.3}_{-100.6}$  \\
$G_{\Theta}$ &$0.46$ & $0.47^{+0.10}_{-0.07}$ \\
\hline
\hline
\end{tabular}
\end{center}
\caption{We demonstrate the recovery of input simulation values from our 
analysis pipeline using the 2D clustering model. The measured values of $D_A$, $H^{-1}$ and $G_{\Theta}$, and their 68\% confidence level uncertainties for 
each realization, agree well with the input simulation values.} 
\label{table:simulation}
\end{table}

%%%%%%%%%%%%%%%%%%%%%%%%%%%%%%%%%%%%%%%%%%%%%%%%%%%%%
\section{Measurements} \label{sec:measure} 

\begin{figure}
\begin{center}
\resizebox{3.4in}{!}{\includegraphics{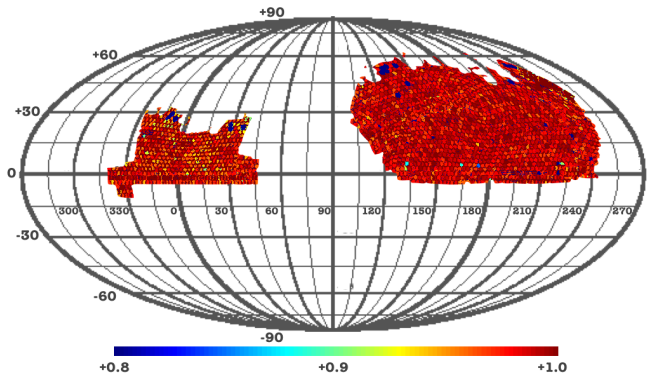}}
\end{center}
\caption{The DR11 CMASS completeness map in Mollweide projection. The North and South patches are centred at RA-Dec locations (185, 25) and (2, 10) respectively.} 
\label{fig:map}
\end{figure}

%\subsection{Data Sample} 
In our analysis we utilise the updated data release (DR11) of the Baryon Oscillation Spectroscopic Survey \citep[BOSS; ][]{2012AJ....144..144B, 2013AJ....145...10D, 2013AJ....146...32S} which is part of the larger Sloan Digital Sky Survey \citep[SDSS; ][]{2000AJ....120.1579Y, 2006AJ....131.2332G} program. From DR11 we focus our analysis on the {\em Constant Stellar Mass Sample} (CMASS) \cite{CMASS}, which contains  690,826 galaxies  and covers the redshift range $z=0.43-0.7$ over a sky area of $\sim$8,500 square degrees. 
The angular coverage of DR11 CMASS is shown in Fig.~\ref{fig:map}. 
The majority of CMASS galaxies are bright, central galaxies (in the halo model framework) and are thus highly biased ($b\sim2$) \citep{2013MNRAS.432..743N}.

%%%%%%%%%%%%%%%%%%%%%%%%%%%%%%%%%%%% 

Each spectroscopically observed galaxy is weighted to account for three 
distinct observational effects: redshift failure, $w_{fail}$; minimum 
variance, $w_{\rm FKP}$\cite{1994ApJ...426...23F}; and angular variation, $w_{sys}$, which accounts for 
airmass dependent seeing effects and stellar contamination. These weights are 
described in more detail in \cite{2012MNRAS.427.3435A} and 
\cite{2012MNRAS.424..564R}. 
The total weight for each galaxy is calculated as the 
product of these weights, i.e., $w_{total}=w_{fail} w_{\rm FKP} w_{sys}$.  
The random catalog points are also weighted but they only include the 
minimum variance FKP weight. 

\begin{figure}
\begin{center}
\resizebox{3.4in}{!}{\includegraphics{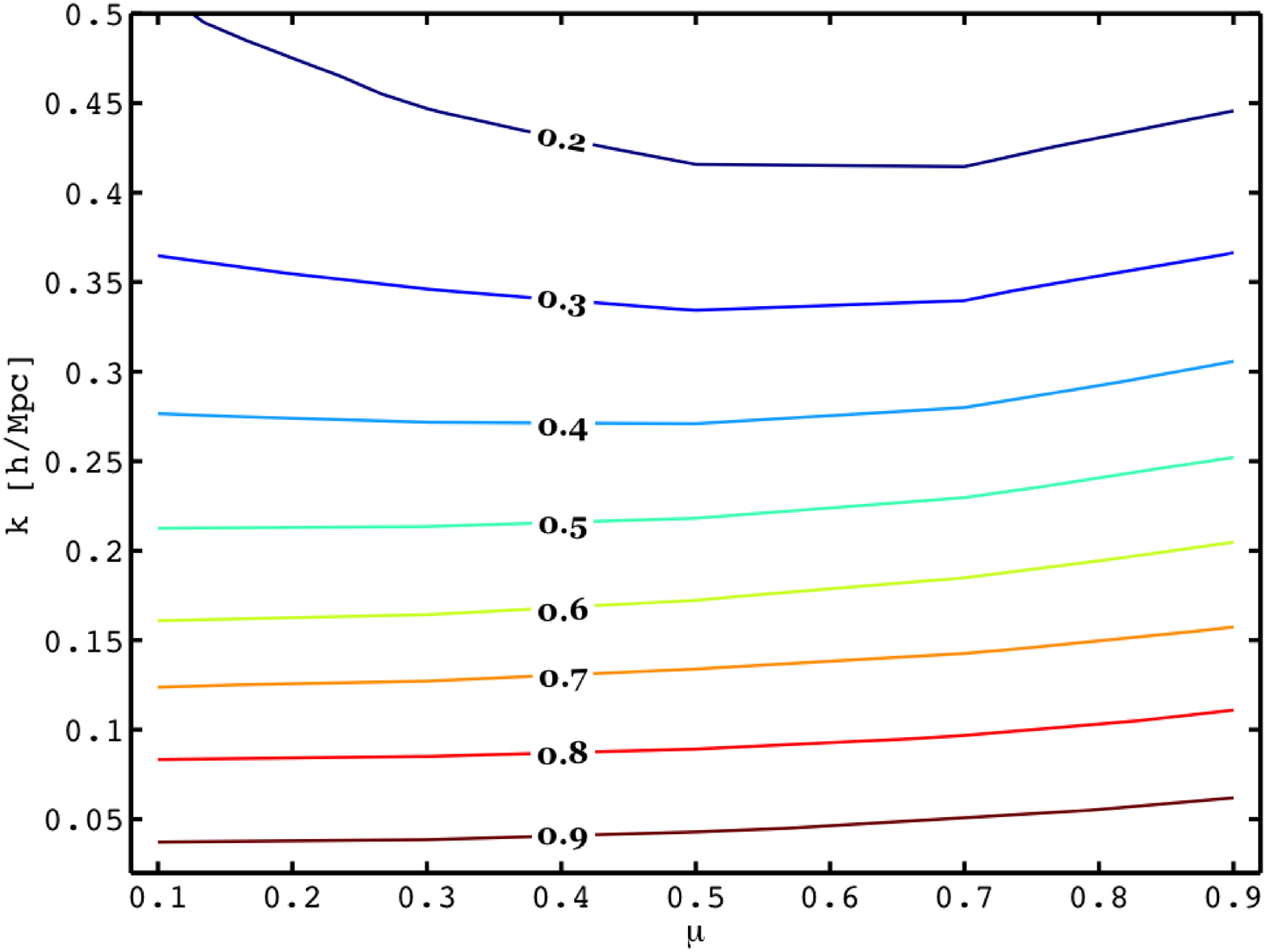}}
\end{center}
\caption{The normalized effective volume $v(k,\mu)$, showing the 
sample variance completeness, is plotted. Values near unity mean that 
new volumes must be surveyed to gain further information on the density 
and velocity fields; lower values 
mean further galaxies can contribute information.} 
\label{fig:veff}
\end{figure}

The CMASS galaxy sample is distributed over the range $0.43<z<0.7$, with an 
effective redshift of  $\zeff=0.57$ 
and an effective volume of  $V_{\rm eff}\sim 6.0\,$Gpc$^3$ calculated as 
\be
V_{\rm eff}=\sum\left(\frac{{n}(z_i)P_0}{1+{n}(z_i)P_0}\right)^2\Delta V(z_i) 
\ , 
\ee
where $\Delta V(z)$ is the volume of a shell at redshift $z$ and 
$P_0=20,000h^{-1}$Mpc. 
To sample the density and velocity fields more fully, one can either 
increase the sampling density (raise $nP$) or survey new volumes. 
Figure~\ref{fig:veff} shows the sample variance completeness of the DR11 
volume for each Fourier mode $k$ and $\mu$, i.e.\ 
\be 
v(k,\mu)=\left(\frac{\bar n P(k,\mu)}{1+\bar n P(k,\mu)}\right)^2 \ , 
\ee 
where $\bar n$ is the average galaxy density within the volume, and the anisotropic power spectrum $P(k,\mu)$ is measured from mock simulations in \cite{2012JCAP...11..014O}. When 
$v$ approaches unity (for high $nP$) then further information on the 
density field at that $k$--$\mu$ can only come from surveying different 
volumes; Fig.~\ref{fig:veff} can thus be thought of as showing the sample 
variance completeness.

%%%%%%%%%%%%%%%%%%%%%%%%%%%%%% 

%\subsection{Measuring the correlation function}\label{sec:2pcf_boss}

We compute the redshift-space 2-dimensional correlation function 
$\xi(\sigma,\pi)$ of the BOSS DR11 galaxy catalog 
using the standard Landy-Szalay estimator~\cite{1993ApJ...412...64L}. 
In the computation of the correlation estimator we use a random point catalogue that constitutes an unclustered but observationally representative sample of the BOSS CMASS survey. The angular points are chosen to reside within the survey geometry with a Monte Carlo acceptance proportional to the RA-Dec sector completeness, and the redshifts are obtained via the random shuffle method of \cite{2012MNRAS.424..564R}. The randoms are also assigned FKP weights, just as for the galaxies. To reduce the statistical variance of the estimator we use $\sim50$ times as many randoms as we have galaxies. 
We perform the coordinate transforms using the fiducial Planck best fit 
flat $\Lambda$CDM cosmological model ($\omega_b=0.022068$, 
$\omega_c=0.12029$, $h=0.67$). 

%%%%%%%%%%%%%%%%%%%%%%%%%%%%%%%%%%%%%%%%% 
\begin{figure}
\begin{center}
\resizebox{3.2in}{!}{\includegraphics{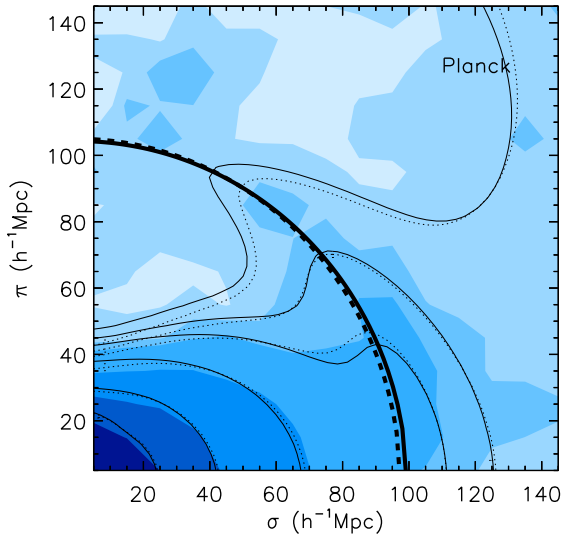}}\hfill 
\end{center}
\caption{The measured (blue filled contours), best fit (thin black), and LCDM-predicted (thin dotted) cases of $\xi(\sigma,\pi)$ are plotted, using the Planck early universe prior. The thick solid 
and dashed circles represent 2D BAO rings from the measurement and the Planck LCDM prediction.}
\label{fig:xi_combined}
\end{figure}

We calculate the correlation function in 225 bins spaced by $10\mpcoh$ 
in the range $0<\sigma,\pi<150\mpcoh$.  
The BOSS sample is naturally separated into North and South samples with 520,805 and 170,021 galaxies respectively, and we measure the anisotropic 
two-point correlation function in the north, south and combined samples. 

%%%%%%%%%%%%%%%%%%%%%%%%%%%%%%%%%%%%%%% 
%\subsection{Covariance matrix} 

Our results depend not only on the correlation functions but also on the errors of these measurements.
Also, different bins of the correlation function can be strongly correlated to each other, therefore 
it is necessary to estimate a covariance matrix to give correct constraints 
on cosmological parameters. 

For this purpose we use {\tt PTHALO} \cite{Manera:2012sc} mock galaxy catalogs, updated to reflect the larger observational area. These catalogs were used for BAO analysis in \cite{Anderson:2014} and have the same survey geometry and number density as the CMASS galaxy sample that was used in our analysis. The mocks catalogs constitute 600 density field realizations  created  using second-order Lagrangian perturbation theory (2LPT). The density and velocity fields created using 2LPT eventually break down as one goes to small scales, but it was confirmed by \cite{Manera:2012sc, Reid:2012sw} that the correlation functions measured from the mock catalogs based on 2LPT match the one measured from the BOSS survey at scales larger than $20\mpcoh$. 
In our analysis below, we will use the data at 
$s\geq 50\mpcoh$ 
and, as we did in \cite{Song:2013ejh}, to be conservative we will additionally remove the data along the line of sight, which is known to deviate from linear theory starting from larger scales than the data perpendicular to the line of sight (see Sec.~\ref{sec:results} below). 
Note that we use these mock catalogs solely to estimate errors of the correlation function measured from the CMASS sample.

For each realization we compute the correlation function as we did for the 
observed catalog and obtain a covariance matrix 
as described in detail in \cite{Linder:2013lza}. 

The measured two point correlation function $\xi(\sigma,\pi)$ is presented as blue filled contours in Fig.~\ref{fig:xi_combined}, with the levels of $(0.2, 0.06, 0.16, 0.005, 0.002, -0.001, -0.006)$ from the inner to outer contours.
The RSD anisotropy is clearly visible, as is the 2D BAO ring at 
$\sqrt{\sigma^2 + \pi^2}\approx 100~\mpcoh$.

%%%%%%%%%%%%%%%%%%%%%%%%%%%%%%%%%%%%%%%%%%%%%%%%%%%
\section{Results} \label{sec:results} 

%%%%%%%%%%%%%%%%%%%%%%%%%%%%%%%%%%%%%%%%%
\subsection{Cosmology from combined maps} 

%%%%%%%%%%%%%%%%%%%
\begin{table}
\begin{center}
\begin{tabular}{c|ccccc}
\hline
\hline
 & Planck LCDM & Combined & North & South\\
\hline
$D_A$       & $936.3$ & $954.9^{+19.8}_{-21.6}$  &$955.1^{+20.8}_{-21.7}$ & $970.4^{+35.8}_{-41.1}$\\
$H^{-1}$   & $2170.8$ & $2159.8^{+136.3}_{-117.8}$  & $2207.2^{+124.9}_{-103.0}$ & $2220.9^{+243.9}_{-358.0}$ \\
$G_b$ &--- & $1.15^{+0.08}_{-0.08}$ & $1.10^{+0.07}_{-0.07}$ & $1.15^{+0.20}_{-0.19}$\\
$G_{\Theta}$ &$0.46$ & $0.41^{+0.09}_{-0.09}$ & $0.34^{+0.08}_{-0.08}$ & $0.54^{+0.16}_{-0.17}$\\
$\sigma_p$ &---&$6.2^{+3.6}_{-3.8}$&$1.5^{+3.3}$& $9.2^{+3.4}_{-4.4}$\\
\hline
\hline
\end{tabular}
\end{center}
\caption{We present the measured values of $D_A\,(\mpcoh)$, $H^{-1}\,(\mpcoh)$, $G_b$, $G_{\Theta}$ and $\sigma_p\, (\hompc)$, and their 68\% CL uncertainties, using the combined data, and the north and the south maps 
separately. Here the cutoffs are $s_{\rm cut}=50\hompc$ and $\sigma_{\rm cut}=40\hompc$.
}\label{tab:measured}
\end{table} 

Information on the late-time cosmological expansion and dynamics are encoded in the 
distances along and transverse to 
the line of sight, and the growth of the density and velocity fields. 
A first  consistency check of $D_A$ and $H^{-1}$ with the Planck 
LCDM model can be ascertained by comparing the two 2D BAO rings in 
Fig.~\ref{fig:xi_combined}. The thick dashed and solid circles represent 
the rings estimated from the Planck LCDM model and the DR11 measurements. The growth functions do not alter the 2D BAO ring, and it is distorted differently by $D_A$ and $H^{-1}$ in the transverse and radial directions. The consistency between the solid and dashed circles means that the measured distances agree well with the Planck LCDM model. 

We quantify this in Table~\ref{tab:measured}, where the measured $D_A$ and 
$H^{-1}$ are within the 68\% confidence limit of the Planck LCDM prediction. 
We discuss the effect of $\scut$ below, and of North vs South maps in the 
next section.

%%%%%%%%%%%%%%%%%%%%%%%%%%%%%%%%%%%%%% 
\begin{figure}
\begin{center}
\resizebox{2.8in}{!}{\includegraphics{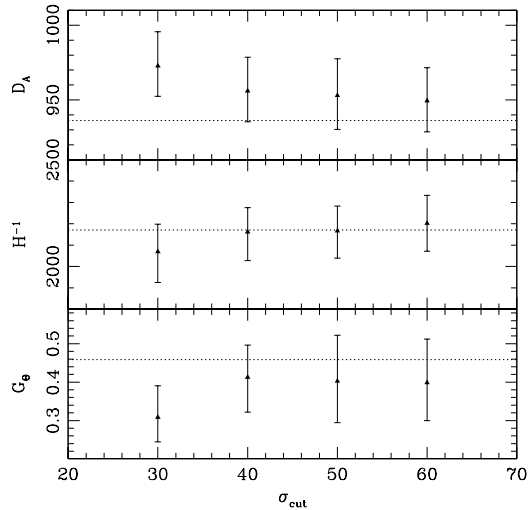}}\hfill 
\end{center}
\caption{The measured values of $D_A$, $H^{-1}$ and $G_{\Theta}$ are shown 
for various $\sigma_{\rm cut}$ from $30\hompc$ and $60\hompc$. The values 
have converged for $\sigma_{\rm cut}\geq 40\hompc$, but inclusion of smaller 
scales biases the answers by $\sim1\sigma$. The dotted lines, representing 
the Planck LCDM predictions, are shown purely for reference; the important 
aspect is convergence (not to any particular value). 
}
\label{fig:measured_combined}
\end{figure}

For the density and velocity growth factors the information comes from 
multiple scales, and especially from the redshift space anisotropy. While 
the signal--to--noise of the inner (higher amplitude) contours of clustering 
is higher, 
the use of {\tt RegPT} to second order is insufficient for accurate 
modelling of $\xi(\sigma,\pi)$ at scales $s<50\hompc$. In particular, the 
cross--spectrum between $\delta$ and $\Theta$ is not perfectly 
cross--correlated. When the cut--off $s<50\hompc$ is applied, the 
constraints on $G_b$ and $G_{\Theta}$ from their distinctive amplification 
of the inner contours~\cite{Song:2010kq} become weaker, but more robust as 
we now see. 

At small scales, if the non--perturbative effect of FoG is underestimated, 
then the residual squeezing can be misinterpreted as a smaller $\gth$. 
We expect the FoG effect to be increasingly important at smaller scales, 
and so these run increasing risk of misestimation. To test this, in 
Fig.~\ref{fig:measured_combined} we show the cosmology results as 
we vary $\sigma_{\rm cut}$ from $30\hompc$ to $60\hompc$. The strongest 
effect is on $\gth$, and indeed inclusion of small scales noticeably lowers 
the measured $G_{\Theta}$. However, for all $\scut\ge 40\hompc$ the 
results have converged and the measured values are insensitive to the exact 
value of $\scut$. This indicates that the approximation for treating FoG should 
not be trusted for $\sigma<40\hompc$, while above this scale our approach is 
robust. This argument holds as well for $D_A$ and $H^{-1}$, though less extremely. 
Considerable caution should be applied to the use of the clustering data 
on small scales.

%%%%%%%%%%%%%%%%%% 
\begin{figure*}
\begin{center}
\resizebox{3.4in}{!}{\includegraphics{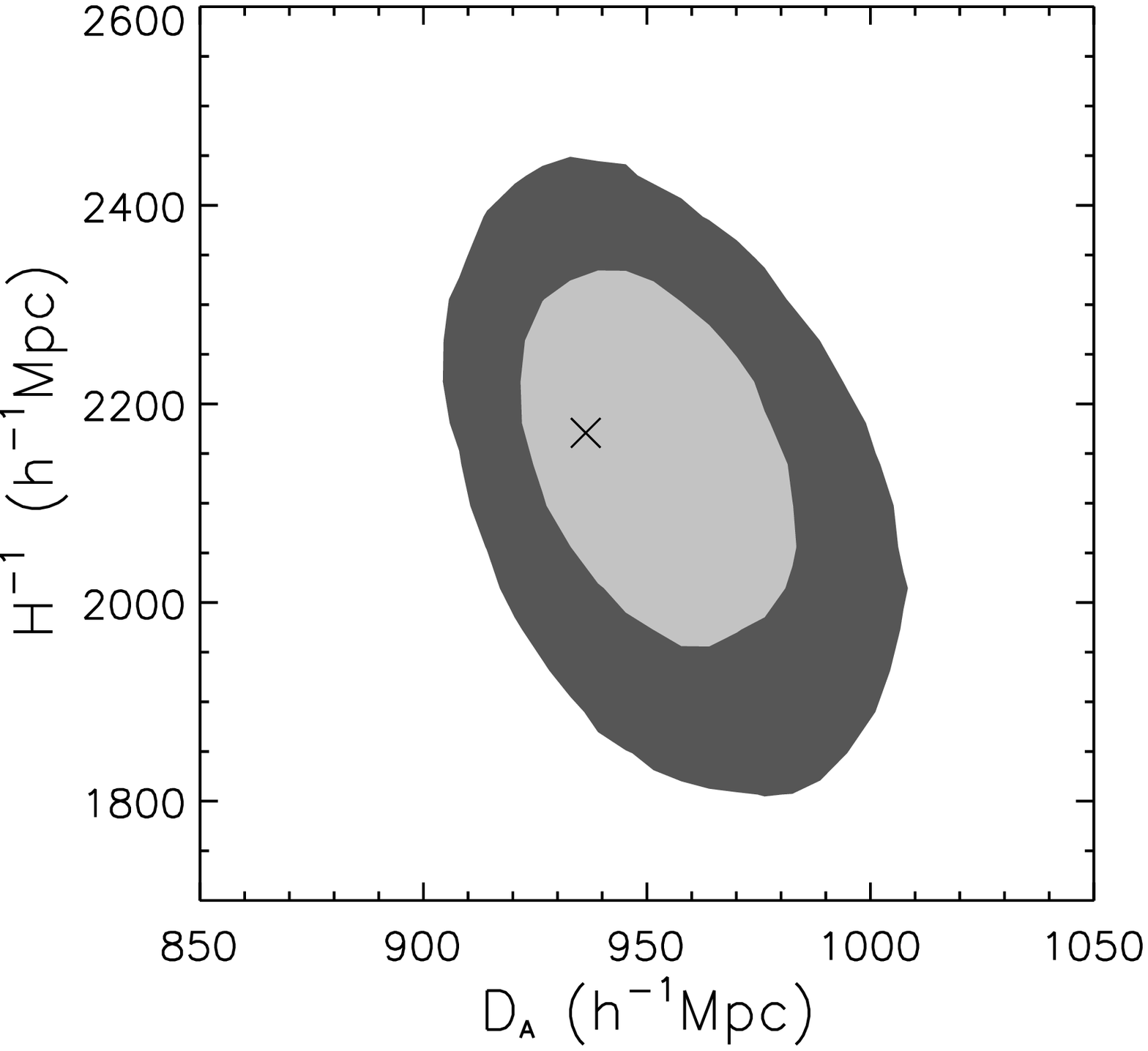}}
\resizebox{3.4in}{!}{\includegraphics{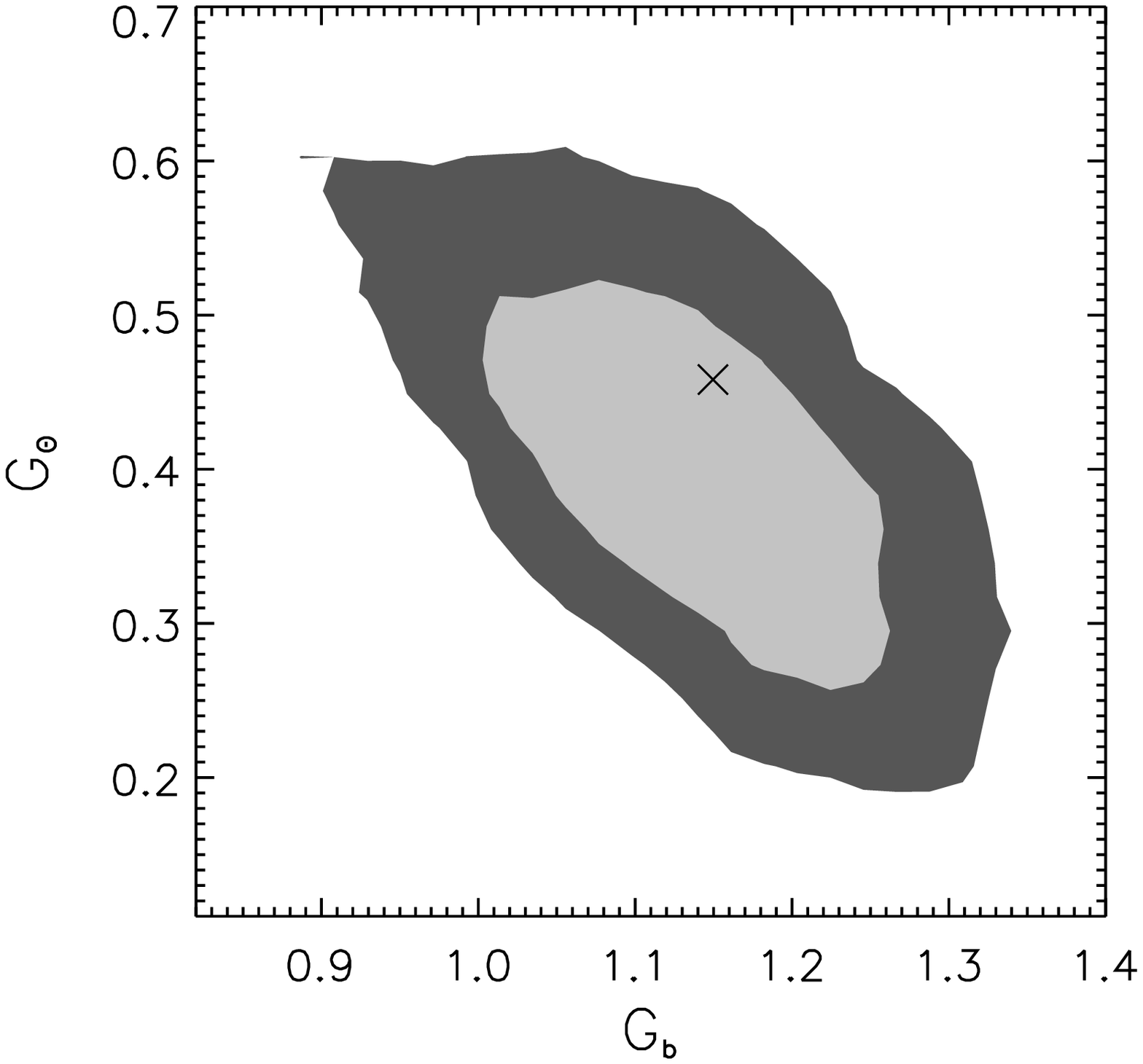}}\\ 
\resizebox{3.4in}{!}{\includegraphics{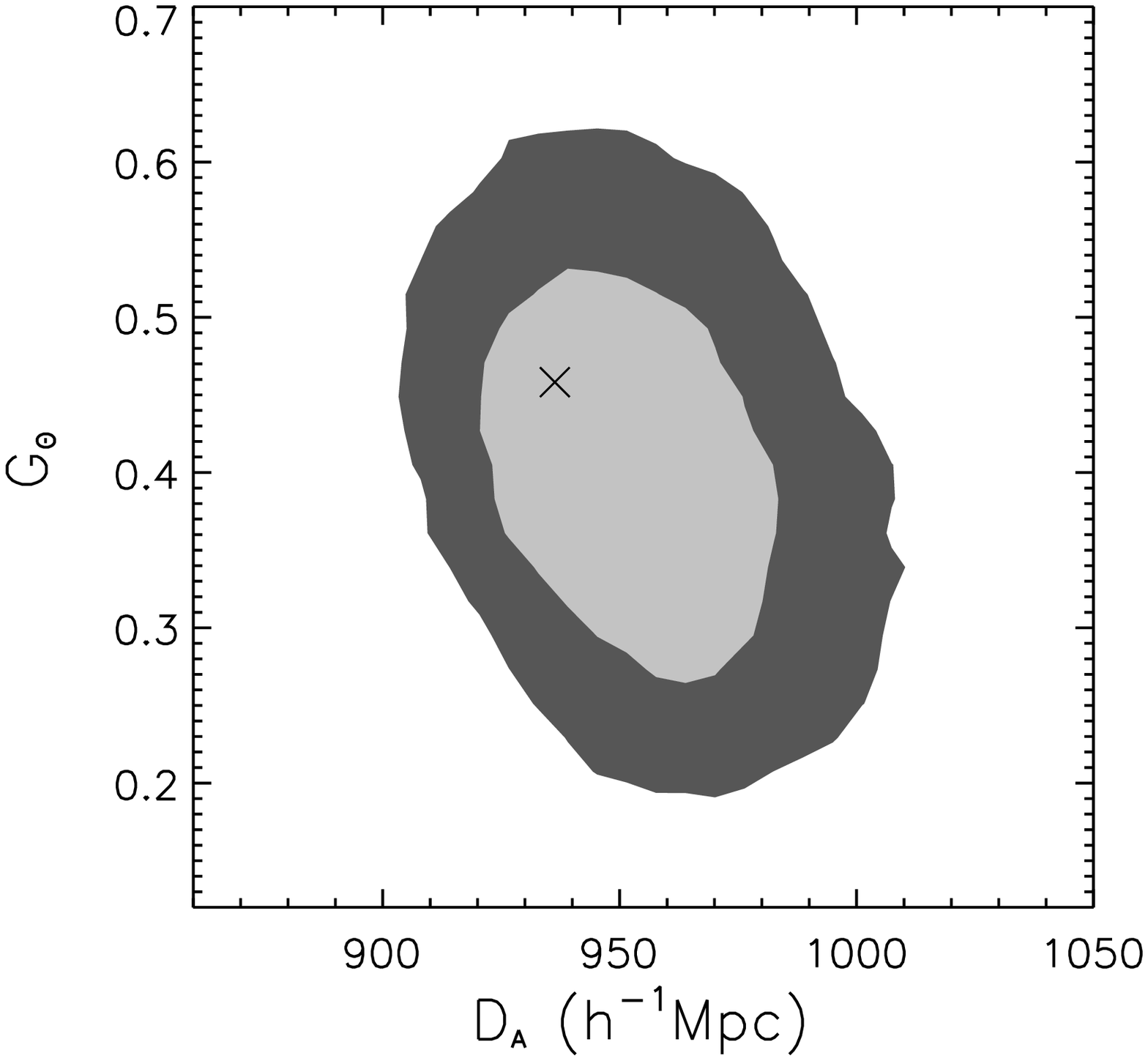}}
\resizebox{3.4in}{!}{\includegraphics{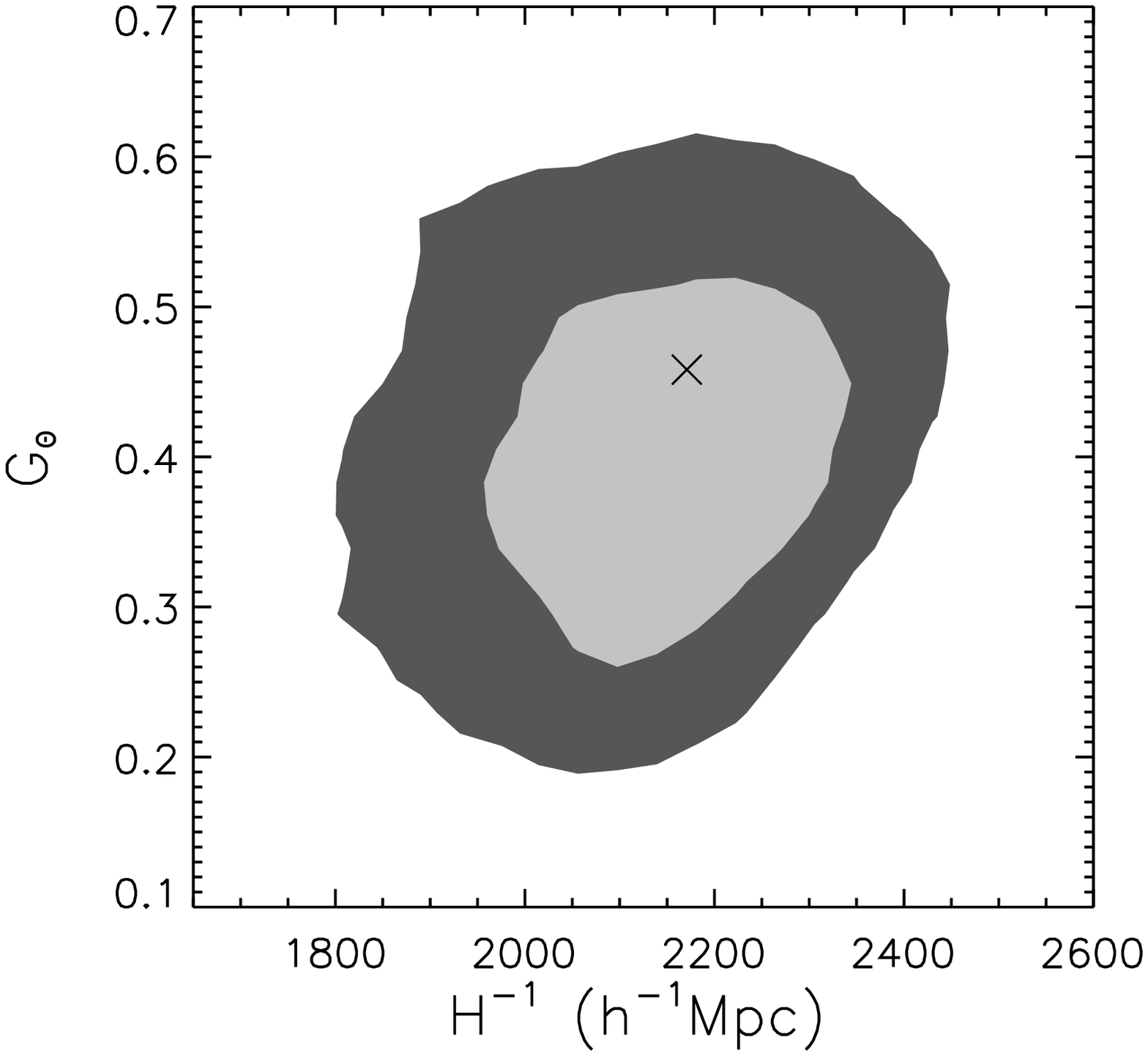}}
\end{center}
\caption{The 2D joint likelihood contours at 68\% and 95\% CL measured for $D_A$, $H^{-1}$, $G_b$ and $G_{\Theta}$ are shown, using  $s_{\rm cut}=50\hompc$ and $\sigma_{\rm cut}=40\hompc$. The fiducial values in the Planck LCDM concordance model are shown by x's (see Table~\ref{tab:measured}). 
} 
\label{fig:2Dparam}
\end{figure*}

The outer contours provide another indicative behavior for $\gth$. When $G_{\Theta}$ varies, the location of peaks on 2D BAO circle moves differently from the variation of $G_b$~\cite{Song:2013ejh}. The peak points run away from the pivot point (roughly where 
$\sigma\approx\pi$) as $G_{\Theta}$ decreases, and move toward it for 
increasing 
$G_{\Theta}$. From Fig.~\ref{fig:xi_combined} we see that around the BAO 
ring the fourth and sixth contours recede away from the Planck LCDM 
prediction, while the fifth contour, which lies close to the pivot point, 
has not moved. 
This implies the measured $G_{\Theta}$ is smaller than the Planck LCDM 
prediction and indeed we find $G_{\Theta}=0.41\pm0.09$ in comparison with 
0.46 predicted by Planck LCDM prediction. This is still within 68\% CL 
however (note though that using a lower $\scut$ moves $\gth$ to even 
smaller values). 

The quantity $G_b$ represents the combination of density field and linear 
bias. As no cosmological model is assumed, the two are not separable. As 
shown in Table~\ref{tab:measured}, $G_b$ is measured to be $1.15\pm0.08$. 
If the Planck LCDM model is true, then the linear bias at $z=0.57$ can be 
estimated to be 
$b=2.0$ with 7\% fractional error (recall that $G_b=bD$). 
This is consistent with what we observe from the simulation. 

The velocity dispersion parameter $\sigma_p$ for FoG is measured to be 
$6.2^{+3.6}_{-3.3}\mpcoh$. The FoG effect is at first order degenerate 
with $G_{\Theta}$; this causes weak constraints on both $\sigma_p$ and 
$G_{\Theta}$.

%%%%%%%%%%%%%%%%%%%%%%%%%%%%%%%%%%%% 
\subsection{Testing the cosmological framework} \label{sec:framework} 

Our model independent analysis allows several consistency tests. 

\begin{enumerate} 

\item In a Friedmann-Robertson-Walker (FRW) cosmology, $D_A$ is formed from 
an integral over $H^{-1}$. 

\item Within general relativity and FRW, the growth rate $\gth$ and expansion rate $H^{-1}$ are tied together. 

\item Neutrino mass suppresses growth, so a measured consistency with Planck 
LCDM (i.e.\ minimal neutrino mass) disfavors higher neutrino mass (or 
requires a conspiracy with enhanced growth from modified gravity -- but this 
would show up in the previous consistency test). 

\end{enumerate} 

We therefore study the joint probability distribution between the measured 
cosmological quantities, e.g.\ the two dimensional likelihood contours of 
$\gth$ vs $H^{-1}$. All such contours are marginalized over the remaining 
quantities. 

In Fig.~\ref{fig:2Dparam} we present two dimensional cosmological parameter 
contours in four different combinations. The x's represent the best values 
of Planck LCDM models; we see that our model independent analysis agrees 
within the 68\% confidence level with the cosmology that assumed LCDM, 
general relativity, and minimal neutrino mass. This holds for all the 
measured distances and growth functions. 

The first panel provides evidence for the FRW consistency relation of the 
background quantities of the distance and expansion rate. The second panel 
involves measurements of the perturbed quantities, from the density and 
velocity fields. As mentioned, $G_b$ is consistent with expectations for 
galaxy bias. The measured $G_{\Theta}$ can be converted to 
$f\sigma_8=0.43\pm0.09$, with the Planck LCDM model predicting 
$f\sigma_8=0.48$, again within 68\% CL. (Note this would not hold if we 
naively included smaller scales where nonlinear modelling is not robust.) 

The third and fourth panels, showing that the joint likelihoods for the 
background and growth quantities are consistent with Planck LCDM, can be 
thought of as a (weak, model independent) test of the general relativity 
criterion. That the fourth panel, showing the $\gth$--$H^{-1}$ likelihood, 
is consistent with Planck LCDM (with minimal neutrino mass), also disfavors 
a larger neutrino mass and its accompanying suppression of growth. If one 
assumed that the background cosmology is truly the Planck LCDM model, then 
the growth measurement could be converted to an estimate of the 
gravitational growth index $\gamma$ \cite{Linder2005} 
or a sum of neutrino masses $\sum m_\nu$. However for both of these the 
uncertainty on $\gth$ is multiplied by a large prefactor so the constraints 
are weak.

%%%%%%%%%%%%%%%%%%%%%%%%%%%%%%%%%%%%%%%%%%%%%%%%%%%%%%%%%%%%%%%%%%%% 
\subsection{Comparison of North vs South} \label{sec:northsouth} 

\begin{figure}
\begin{center}
\resizebox{2.8in}{!}{\includegraphics{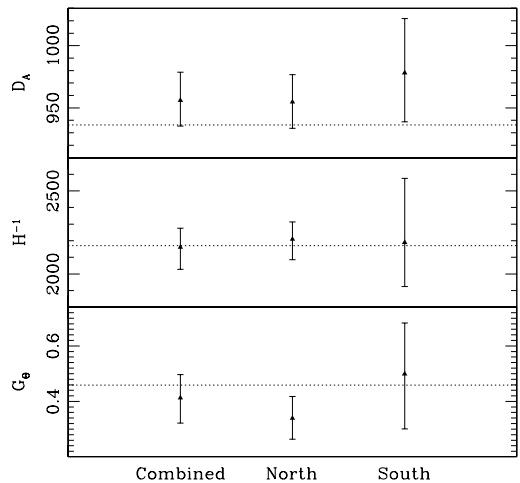}}\hfill 
\end{center}
\caption{The measured $D_A$, $H^{-1}$ and $G_\Theta$ are presented from 
the top to the bottom panels. Each one shows the results from the combined, 
northern, and southern skies, from left to right. 
The dotted lines represent the Planck LCDM predictions. 
}
\label{fig:errors_cns}
\end{figure}

Another interesting check involves a comparison of the estimated cosmological 
quantities using only the northern or southern hemisphere sky. The median RA 
and Dec of each patch is $(185,25)$ and $(2,10)$ respectively (see 
Fig.~\ref{fig:map}), so the 
centers of these two disjoint sky patches are separated by 145 degrees on 
the sky. The effective volume of the North is 4.5 Gpc$^3$ and that of the 
South is 1.5 Gpc$^3$. 

Table~\ref{tab:measured} breaks down the cosmological results by hemisphere. 
All quantities are consistent within 68\% CL; nevertheless, there are some 
interesting patterns worthwhile keeping an eye on as the data improves and 
the error bars shrink. 

In the top and middle panels of Fig.~\ref{fig:errors_cns} we present the 
measured $D_A$ and $H^{-1}$ for North and South separately, and the full 
survey combination. North and South agree with each other and with Planck 
LCDM predictions. The detailed numbers are shown in Table~\ref{tab:measured}. 
Note that the median measured $H^{-1}$ from the combined map does not lie 
between the North and South measured values; this occurs due to the 
non-Gaussian probability distribution for the measured $H^{-1}$ in the 
South -- the mode value is 2116 $h^{-1}\,$Mpc. 
%\scr{YSS: I investigate the LnL plot of 1/H. For north and the combined, it is pretty close to Gaussian, i.e. the most frequent point is nearly same with mid-point of probability. But for the south, it is significantly non-gaussian. The most peak point of 2116 is not the middle point of probability. I report the value at the middle point of probability.}

For the measured coherent motion $\gth$ in the bottom panel, the North is 
somewhat inconsistent ($\sim1.5\sigma$) with the Planck LCDM prediction. From 
Table~\ref{tab:measured}, in the North $G_{\Theta}=0.34^{+0.08}_{-0.08}$ 
and in the South $\gth=0.54^{+0.16}_{-0.17}$, while the LCDM fiducial 
has $G_{\Theta}=0.46$. The central values of North and South are noticeably 
different, though due to the large uncertainty from the small effective 
volume in the South this cannot be said to be statistically significant. 

One might speculate about North-South anisotropy but this is disfavored due 
to the consistency of the measured $H^{-1}$ values. We have also checked 
that $z_{\rm eff}$ is consistent between North and South, at the 0.05\% 
level. Another possibility is 
inhomogeneity at the perturbation level, for example an anisotropic 
stress \cite{DeFelice:2013bxa,Gumrukcuoglu:2013nqa}. 
With a quadrupole dependence, this would not have an effect if the North 
and South areas were 180$^\circ$ apart, but could have a component as they 
are separated by 145$^\circ$. Also note that the measured velocity 
dispersion $\sigma_p$ is different in North and South, which could support 
this. Alternately, the covariance between $\sigma_p$ and $\gth$ is such that 
high $\sigma_p$ can damp the excess velocity growth of high $\gth$, so that 
these (and the low $\sigma_p$, low $\gth$ case for the North) lie along the 
degeneracy direction with LCDM.

%%%%%%%%%%%%%%%%%%%%%%%%%%%%%%%%% 
\section{Conclusions} \label{sec:concl} 

We have carried out an analysis within a framework independent of the 
cosmological model, i.e.\ the specific energy density components such as 
dark energy or curvature. This uses the BOSS DR11 dataset that measures 
galaxy clustering over the largest volume yet surveyed with an effective 
redshift of $z=0.57$. We measure the angular distance $D_A$, expansion 
rate $H^{-1}$ (from the radial distance information), and velocity growth 
rate $\gth$; all are consistent with the Planck LCDM prediction. 

These are multiple, model independent tests of LCDM since the implications 
of each parameter is different. The measured $D_A$ is insensitive to 
uncertainties from contamination along the line of sight, but can be 
affected by the assumption of coherent (scale-independent) galaxy bias 
on the scales used, due to the degeneracy between $D_A$ and $G_b$. 
The value of the galaxy bias $b$ we derive is also consistent with other 
measurements. 
For accurate measurement of $H^{-1}$, the radial dependence of 
$\xi(\sigma,\pi)$ should be modelled robustly. 

For $\gth$, 
measured coherent motions are degenerate with the FoG effect which is 
problematic to model. If there is residual contamination from inaccurate 
modelling of the FoG effect, the coherent motions and hence velocity growth 
rate $\gth$ or $f\sigma_8$ are underestimated. Since the theoretical model 
calibrated from simulations becomes increasingly inaccurate on small scales, 
we carefully examine the dependence of the results on the small scale cutoff 
in the measurements used. We find that convergence is achieved for 
$\scut\ge40\,\hompc$, with bias arising if smaller scales are included -- 
an important caution. Our measurement corresponds to 
$f\sigma_8=0.43\pm0.09$, with the Planck LCDM prediction of 
$f\sigma_8=0.48$. 

Considering the joint likelihood of these cosmological quantities, we 
find consistency with the Planck LCDM model at 68\% CL. By comparing the 
quantities to each other, we can make three general consistency tests of 
the cosmological framework. We check consistency with the FRW framework, 
with general relativity, and with minimal neutrino mass and find that all 
are within 68\% CL.  

Comparing the galaxy measurements from the North and South sky samples 
separately, we continue to find consistency with Planck LCDM. Here the 
reduced effective volume makes the error bars larger, but there are slight 
discrepancies worth testing with future data from larger surveys such as 
DESI or LSST. For example, the growth rate in the North has $\gth=0.34$ 
($\sim1.5\sigma$ from Planck LCDM) while the South has $\gth=0.54$ (though 
again the error bar is so large that these are consistent within 68\% CL). 
One speculative explanation is the presence of anisotropic stress (affecting 
the perturbations while keeping the background quantities consistent), 
but covariance between $\gth$ and $\sigma_p$ (also measured to be different 
in North and South) is another possibility. Again, upcoming larger sky 
surveys will be valuable in testing cosmology in different directions. 

%%%%%%%%%%%%%%%%%%%%%
\acknowledgements 

Numerical calculations were performed by using a high performance computing cluster in the Korea Astronomy and Space Science Institute and we also thank the Korea Institute for Advanced Study for providing computing resources (KIAS Center for Advanced Computation Linux Cluster System). EL was supported in part by US DOE grant DE-SC-0007867 and Contract No. DE-AC02-05CH11231. We thank Marc Manera for providing the mock simulations and Shinji Mukohyama for helpful comments on anisotropic stress.

Funding for SDSS-III has been provided by the Alfred P. Sloan Foundation, the Participating Institutions, the National Science Foundation, and the U.S. Department of Energy Office of Science. The SDSS-III web site is http://www.sdss3.org/. 

SDSS-III is managed by the Astrophysical Research Consortium for the Participating Institutions of the SDSS-III Collaboration including the University of Arizona, the Brazilian Participation Group, Brookhaven National Laboratory, Carnegie Mellon University, University of Florida, the French Participation Group, the German Participation Group, Harvard University, the Instituto de Astrofisica de Canarias, the Michigan State/Notre Dame/JINA Participation Group, Johns Hopkins University, Lawrence Berkeley National Laboratory, Max Planck Institute for Astrophysics, Max Planck Institute for Extraterrestrial Physics, New Mexico State University, New York University, Ohio State University, Pennsylvania State University, University of Portsmouth, Princeton University, the Spanish Participation Group, University of Tokyo, University of Utah, Vanderbilt University, University of Virginia, University of Washington, and Yale University. 

%%%%%%%%%%%%%%%%%%%%%%%%%%%%%%%%% 

\end{document}